\documentclass[twocolumn]{aastex631} %

\usepackage{booktabs}
\usepackage{CJK}
\usepackage{amsmath}

\usepackage{color}
\usepackage{url}
\usepackage{hyperref}
\usepackage{xspace}
\usepackage{soul}
\usepackage{multirow}
\usepackage{threeparttable}

\newcommand{\psrtar}{SRGA\,J144459.2--604207}

\newcommand{\nicer}{{NICER}\xspace}  

\newcommand{\swift}{{Swift}\xspace}
\newcommand{\Integ}{{INTEGRAL}\xspace}  
\newcommand{\maxi}{{MAXI}\xspace}  

\newcommand{\hxmt}{{\it Insight}-HXMT}\xspace

\def\be{\begin{equation}} 
\def\ee{\end{equation}}

\begin{document}
\title{A Comprehensive Study of Type I (thermonuclear) Bursts in the New Transient SRGA J144459.2--604207}

\correspondingauthor{Zhaosheng Li, Yuanyue Pan}
\email{lizhaosheng@xtu.edu.cn, panyy@xtu.edu.cn}

\author{Tao Fu}
\affiliation{Key Laboratory of Stars and Interstellar Medium, Xiangtan University, Xiangtan 411105, Hunan, People's Republic of China}

\author[0000-0003-2310-8105]{Zhaosheng Li}
\affiliation{Key Laboratory of Stars and Interstellar Medium, Xiangtan University, Xiangtan 411105, Hunan, People's Republic of China}

\author{Yuanyue Pan}
\affiliation{Key Laboratory of Stars and Interstellar Medium, Xiangtan University, Xiangtan 411105, Hunan, People's Republic of China}

\author[0000-0001-9599-7285]{Long Ji}
\affiliation{School of Physics and Astronomy, Sun Yat-sen University, Zhuhai 519082, People's Republic of China}

\author[0000-0001-8768-3294]{Yupeng Chen}
\affiliation{Key Laboratory of Particle Astrophysics, Institute of High Energy Physics, Chinese Academy of Sciences, 19B Yuquan Road,
Beijing 100049, People's Republic of China}

\author[0000-0002-7889-6586]{Lucien Kuiper}
\affiliation{SRON-Netherlands Institute for Space Research, Niels Bohrweg 4, 2333 CA, Leiden, The Netherlands}

\author[0000-0002-6558-5121]{Duncan K. Galloway}
\affiliation{School of Physics and Astronomy, Monash University, Victoria 3800, Australia}
\affiliation{Institute for Globally Distributed Open Research and Education (IGDORE), Sweden}

\author[0000-0003-3095-6065]{Maurizio Falanga}
\affiliation{International Space Science Institute (ISSI), Hallerstrasse 6, 3012 Bern, Switzerland}
\affiliation{Physikalisches Institut, University of Bern, Sidlerstrasse 5, 3012 Bern, Switzerland}

\author[0000-0002-9042-3044]{Renxin Xu}
\affiliation{Department of Astronomy, School of Physics, Peking University, Beijing 100871, People's Republic of China}
\affiliation{Kavli Institute for Astronomy and Astrophysics, Peking University, Beijing 100871, People's Republic of China}

\author{Xiaobo Li}
\affiliation{Key Laboratory of Particle Astrophysics, Institute of High Energy Physics, Chinese Academy of Sciences, 19B Yuquan Road,
Beijing 100049, People's Republic of China}

\author[0000-0002-3776-4536]{Mingyu Ge}
\affiliation{Key Laboratory of Particle Astrophysics, Institute of High Energy Physics, Chinese Academy of Sciences, 19B Yuquan Road,
Beijing 100049, People's Republic of China}

\author[0000-0003-0274-3396]{L. M. Song}
\affiliation{Key Laboratory of Particle Astrophysics, Institute of High Energy Physics, Chinese Academy of Sciences, 19B Yuquan Road,
Beijing 100049, People's Republic of China}

\author{Shu Zhang}
\affiliation{Key Laboratory of Particle Astrophysics, Institute of High Energy Physics, Chinese Academy of Sciences, 19B Yuquan Road,
Beijing 100049, People's Republic of China}

\author[0000-0001-5586-1017]{Shuang-Nan Zhang}
\affiliation{Key Laboratory of Particle Astrophysics, Institute of High Energy Physics, Chinese Academy of Sciences, 19B Yuquan Road,
Beijing 100049, People's Republic of China}

\begin{abstract}

We report an analysis of {\it Insight}-HXMT observations of the newly discovered accreting millisecond pulsar SRGA J144459.2--604207. During the outburst, detected in 2024 February by SRG/ART-XC, the broadband persistent spectrum was well fitted by an absorbed Comptonization model. We detected 60 type I X-ray bursts in the {\it Insight}-HXMT medium energy (ME) data, and 37 were also detected with the low-energy (LE) telescope. By superimposing the {\it Insight}-HXMT/LE/ME/HE light curves of 37 bursts with similar profiles and intensities, we measured a deficit of X-rays in the 40–70 keV energy band. By analyzing
the time-resolved X-ray burst spectra, we determine the mean ratio of persistent to burst flux of $\alpha=71\pm7$.
We estimate the average hydrogen mass fraction in the fuel at ignition, as $\overline{X} =0.342\pm0.033$, and constrain the burst fuel composition as $X_0\lesssim0.4$. We
found that 14 out of 60 X-ray bursts exhibited photospheric expansion, and thus we estimated the distance to
the source as $10.0\pm0.71~\rm{kpc}$. Combined with {\it IXPE} observations, the burst recurrence time increased from 1.55 to 8 hr as the local mass accretion rate decreased, which can be described as $\Delta T_{\rm rec}\sim \dot{m}^{-0.91\pm0.02}$.
\end{abstract}

\section{Introduction}
\label{sec:intro}
Type I X-ray bursts arise from unstable thermonuclear burning that occurs in low-mass X-ray binary (LMXB) systems hosting a neutron star (NS) and a low-mass companion ($ M\lesssim1 M_{\odot}$; \citealt{1992apa..book.....F}). For detailed reviews, see \citet{1993SSRv...62..223L,2006csxs.book.....L}, and \citet{2021ASSL..461..209G}. At lower local mass accretion rates ($\dot{m} \lesssim 10\%$ of the local Eddington accretion rate), typically helium-rich bursts are rapidly and powerfully triggered by the triple-alpha process, reaching a peak with $\sim 1\text{--}2~\rm{s}$ and lasting $\sim 10\text{--}20~\rm{s}$. For these events, the helium comes either from the helium-rich donor or from the stable and (sometimes) complete burning of hydrogen. The burst decay time is attributed to the cooling of the NS photosphere, which leads to a gradual softening of the burst spectrum. For NS LMXBs with high mass accretion rates, type I X-ray bursts typically occur in mixed hydrogen and helium environments powered by unstable helium ignition, where the hydrogen is accreted faster than consumed through a hot CNO cycle limited by $\beta$-decays. Consequently, these bursts exhibit longer rise and decay time \citep[see, e.g.,][for detailed discussions]{1993SSRv...62..223L,2006csxs.book.....L}. The burst recurrence time ($\Delta T_{\text{rec}}$) usually ranges from hours to days, depending on the accretion rate and the composition of the accreted material \citep[see, e.g.,][]{Galloway06}. The relation, $\Delta T_{\text{rec}}\sim\dot{m}^{-1}$, has been found in systems such as the ``clocked'' bursters GS~1826--24 and 1RXS~J180408.9--342058, and two accreting millisecond X-ray pulsars (AMXPs) SAX J1748.9--2021 and MAXI J1816--195 \citep{2004ApJ...601..466G, 2019ApJ...870...64G, 2018A&A...620A.114L, 2024A&A...689A..47W}. 

The X-ray burst spectrum can be described in terms of blackbody radiation with temperature of $kT_{\rm bb}\approx0.5\text{--}3~\rm{keV}$. From the time-resolved burst spectral studies, the flux of type I X-ray bursts can reach the Eddington limit, and the radiation pressure exceeds the gravitational potential on the NS surface, leading to photosphere radius expansion \citep[PRE;][]{1993SSRv...62..223L}. PRE bursts are usually used as standard candles to determine the distance to the source phenomenologically \citep{2003A&A...399..663K}. The interactions between burst radiation and the surrounding accretion disk have been observed as the enhancement of persistent accretion due to the Poynting-Robterson drag \citep{2004ApJ...602L.105B,2013ApJ...772...94W,2014ApJ...797L..23K,2015ApJ...801...60W,Zhao22,Lu23}, the deficient hard X-ray emission due to corona cooling, and the accretion disk reflection \citep{2012ApJ...752L..34C,2013MNRAS.432.2773J,2018ApJ...864L..30C,2019JHEAp..24...23C,Lu24,2024ApJ...976...44S, Yu24}.

SRGA J144459.2--604207 was discovered on 2024 February 21 by the Mikhail Pavlinsky ART-XC telescope survey mission \citep{2024A&A...690A.353M}. Follow-up observations revealed 
the source
is a new AMXP with a spin period of 447.9~Hz, and an orbital period of $\sim 5.2~\rm{hr}$ \citep{2024ATel16548....1L,2024ApJ...968L...7N}. Several X-ray instruments have detected many X-ray bursts from \psrtar, i.e., \nicer\ (5 bursts; \citealt{2024arXiv240800608P}), the \swift X-ray Telescope (1 burst; \citealt{2024ATel16471....1C}), the \maxi Gas Slit Camera (1 burst; \citealt{2024ATel16483....1N}), ART-XC (19 bursts; \citealt{2024A&A...690A.353M}), XMM-Newton (13 bursts; \citealt{2024arXiv240800608P}), NuSTAR (23 bursts; \citealt{2024arXiv240800608P}), NinjaSat (12 bursts and 2 bursts are the same as bursts \#36 and \#40 in \hxmt; \citealt{takeda2024ninjasatmonitoringtypeixray}), Chandra (1 burst; \citealt{2024ATel16510....1I}), \Integ\ \citep{2024ATel16493....1S,2024ATel16507....1S}, and {\it IXPE} (52 bursts and 18 of them also in the Insight Hard X-ray Modulated Telescope (\hxmt)\ burst sample; \citealt{2024arXiv240800608P}). These observations show that the burst recurrence time varied between 1.6 and 10~hr as the persistent count rate decreased.
The {\it IXPE} observations found a relation between the recurrence and the persistent count rate $t_{\rm rec}\sim C^{-0.8}$ \citep{2024arXiv240800608P}.

In this paper, we report the X-ray burst properties of the AMXP SRGA J144459.2--604207 using the \hxmt \ observations. In Sect.~\ref{sec:observation}, we describe the data reduction procedure of \hxmt. In Sect.~\ref{sec:3}, we introduce the characteristics of X-ray burst light curves, the persistent emission, and time-resolved spectra for the X-ray bursts. 
We discuss the results in Sect.~\ref{sec:4}.

\section{Observation} \label{sec:observation}

\hxmt\ \citep{2020SCPMA..6349502Z} has three collimating telescopes, the low-energy (LE) X-ray telescope \citep{2020JHEAp..27...24L}, the medium-energy (ME) X-ray telescope \citep{2020JHEAp..27...44G}, and the high-energy (HE) X-ray telescope \citep{2020SCPMA..6349503L}, which have effective areas of 400, 900, and 5000~cm$^2$, respectively, covering the broadband energy range of 1--250 keV. After \psrtar\ was confirmed as an AMXP, we proposed ToO observations using \hxmt\ (PI: Zhaosheng Li). \hxmt\ conducted 53 observations of SRGA J144459.2--604207 between 2024 February 23 and March 15, with a total exposure time of 691~ks. We processed and analyzed the data using the \hxmt\ Data Analysis Software v2.06. The light curve, source, and background spectra for LE, ME, and HE were extracted using the tool, \texttt{lepipeline}, \texttt{mepipeline}, and \texttt{hepipeline}, respectively. We screened the data according to the standard criteria: the offset angle from the pointing direction $<0.^{\!\!\circ}04$; the pointing direction above Earth $>10^\circ$; and the cutoff rigidity value $>8$~GV. The net exposures obtained from LE, ME, and HE were 104, 123, and 129~ks, respectively.

\begin{figure}
    \centering
    \centerline{\includegraphics[width=1\linewidth]{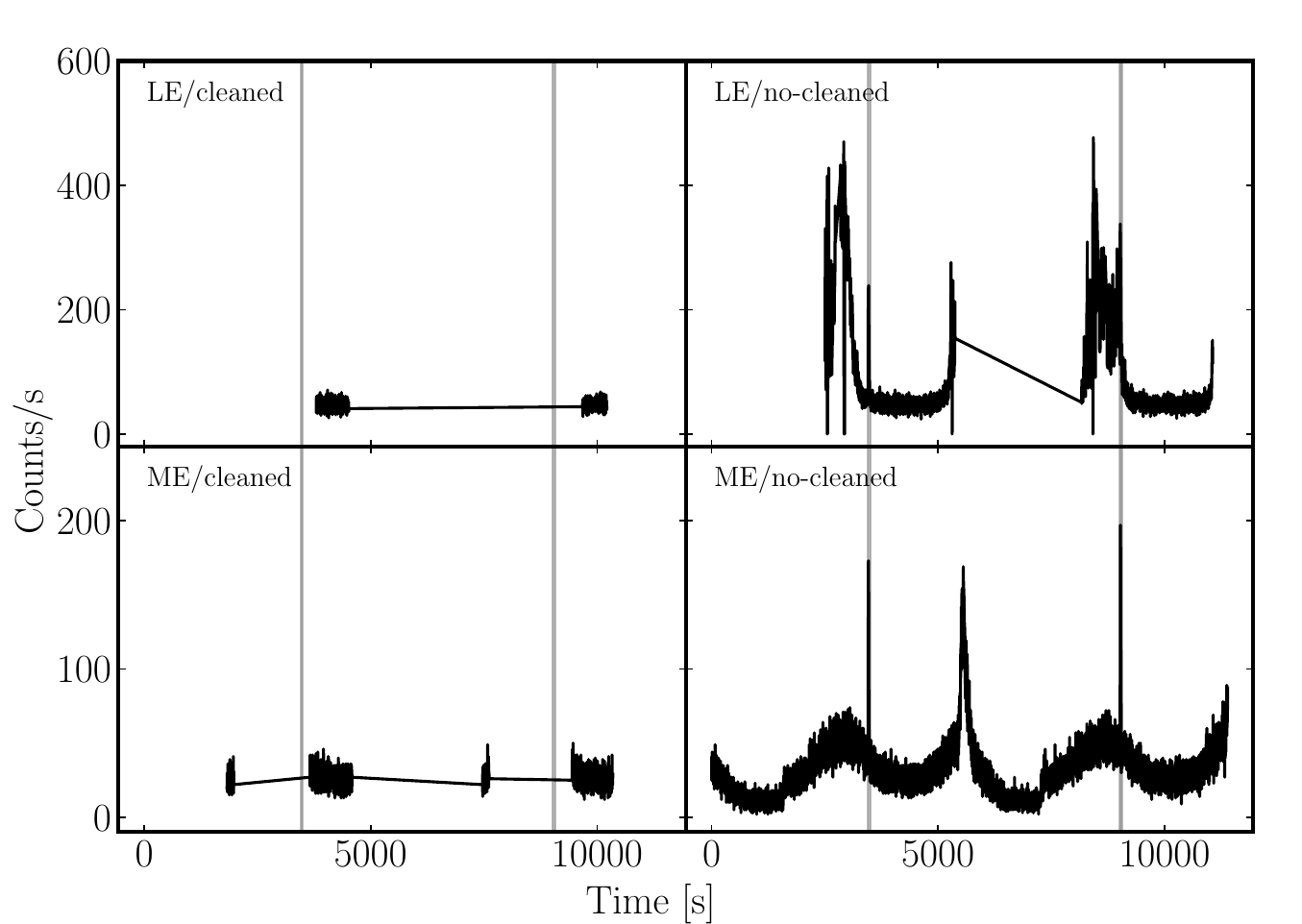}}
    \caption{Light curves from Obs. id. ID P061437300103. In the left panels, the light curves are produced via the standard processing from LE and ME, respectively. In the right panels, the light curves are generated without GTI filtering, and bursts $\#4$ and $\#5$ are detected both by LE (top right panel) and ME (bottom right panel). The gray lines represent the onset time of two bursts.}
    \label{fig:1}
\end{figure}

We note that the default good time interval (GTI) selection criteria for LE and ME are very conservative due to a light leak \citep{2020JHEAp..27...24L}, and many X-ray bursts were missed. We followed the same procedure used for observations of MAXI~J1816--195 \citep{2022ApJ...936L..21C} to extract the light curves and spectra without filtering on GTIs, to obtain a complete sample of bursts from SRGA J144459.2--604207. The light curves of LE, ME, and HE data are extracted with a time bin of 1 s. In Figure \ref{fig:1}, we show the effect of such a treatment as an example, in which the Obs. id. P061437300103 contains bursts \#4 and \#5. These two bursts were missed in the LE and ME light curves generated by the standard GTI filtering criteria. When \hxmt\ passed through the South Atlantic Anomaly, burst-like fluctuations that could be caused by sharp changes in the background were excluded. 

We also processed the {\it IXPE} data of \psrtar\ \citep[see also][]{2024arXiv240800608P}. For each detector unit, the 1-s binned light curves were extracted from a 100'' circle region around the source position. We identified the onset time of 52 type I X-ray bursts. The pre-burst count rates of these bursts were calculated, which were used to convert to the persistent flux (see Sect.~\ref{subsec:4.3}).

\section{Results} \label{sec:3}

\begin{figure*}
    \centerline{\includegraphics[width=1\linewidth]{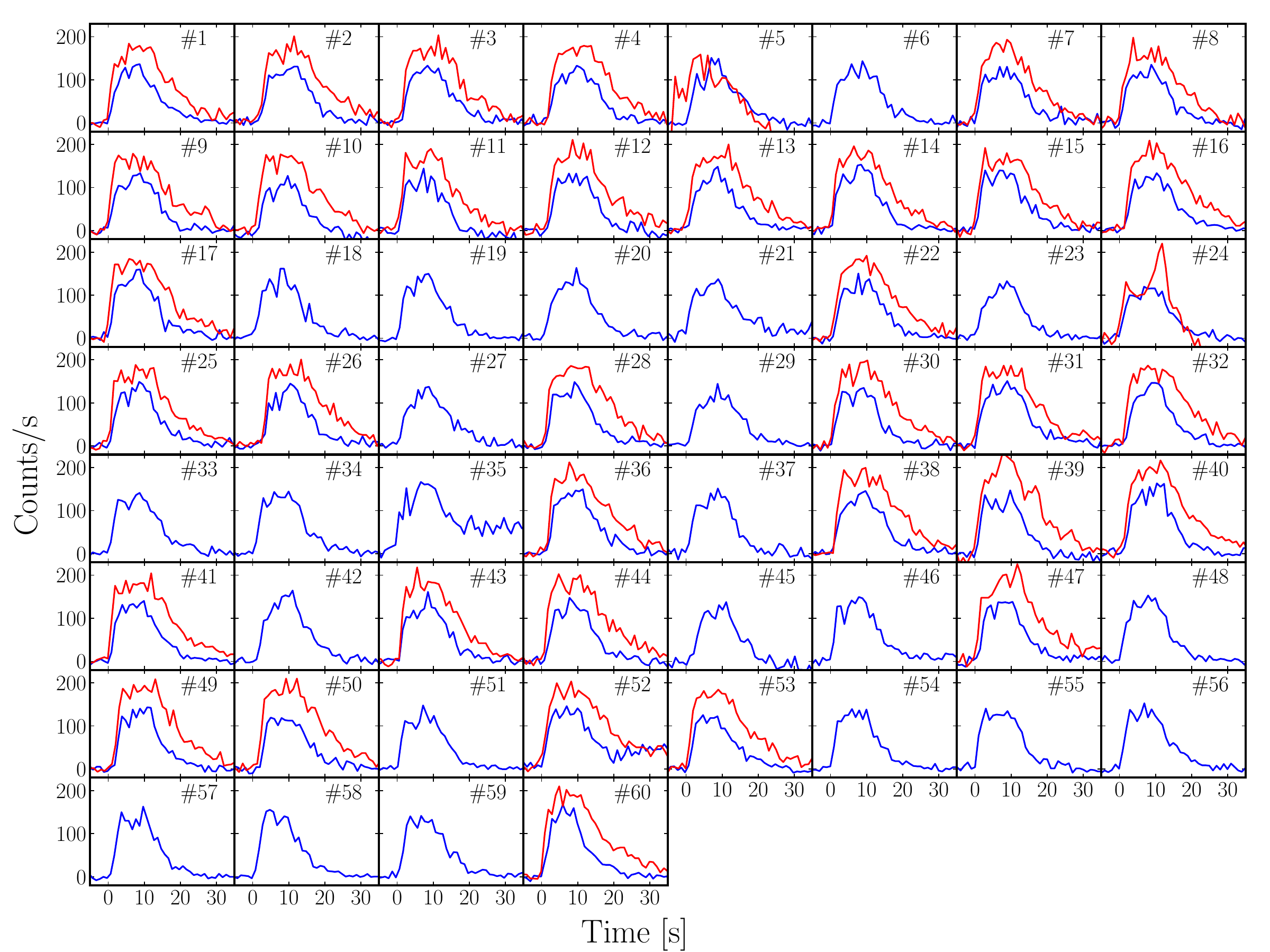}}
    \caption{Light curve of the 60 X-ray bursts from SRGA J144459.2--604207 observed with \hxmt. The red and blue lines represent the light curves of the LE (2--10~keV) and ME (10--35~keV) bursts, respectively. The persistent LE and ME count rates are subtracted independently. Light curves were relative to the burst onset time.}
    \label{fig:2}
\end{figure*}

\subsection{Burst light curves} \label{subsec:3.1}
 We found 60 type I X-ray bursts in the ME data and 37 of them were also detected in the LE data. No bursts were apparent in the HE data, covering 25--250~keV, likely because the burst spectra were too soft. 
From 1~s binned ME light curves, we defined a time interval between 70 and 20~s before the burst peak as the pre-burst emission. At the end of the pre-burst, we defined the time when the count rate was 105\% of the average count rate of the pre-burst emission as the burst onset time. We found that the ME and LE bursts had almost the same onset time, except for burst \#5, for which the LE light curve was $\sim2$~s ahead of ME. We subtracted the average pre-burst count rate for both LE and ME burst light curves, as shown in Figure~\ref{fig:2}. The profiles of the LE and ME light curves for 60 bursts were compared and it was found that the ME light curve profiles in the 10--35~keV range were quite similar. The burst light curves reached a rough plateau with peak count rates of $\sim 130~\rm {cts~s^{-1}}$ and duration of 7--10~s after a fast rise time of 2--3~s, followed by an exponential decay to the pre-burst level. The LE burst light curves had a higher peak count rate, $\sim 180~\rm{cts~s^{-1}}$, than the ME bursts, but the profiles are otherwise quite similar in shape.

We defined the recurrence time as the duration between the burst onset of two successive bursts by incorporating the bursts from {\it IXPE}, NinjaSat, and \Integ, and list the values in Table~\ref{Table:2}. 
Some bursts have recurrence times that are significantly longer than others immediately before or after, which implies that some bursts may have been missed due to gaps in the data. We obtained the true recurrence time by dividing the observed recurrence time by the integer $N+1$, where $N$ is the number of missed bursts. For instance, the $\Delta T_{\rm rec}$ for burst \#41 is 4.49~hr, and by comparing with bursts \#40 and \#42, it is suspected that one burst might be missed. From the {\it IXPE} burst sample, there was a burst on MJD~60368.10542, which perfectly filled the gap between bursts \#40 and \#41, confirming the reliability of our method for obtaining the true recurrence time. Please note that this method is the key assumption to estimate the recurrence time, which is only valid for the sources with repeating bursts. We found that as the persistent emission decreased, the recurrence time for the bursts observed with {\it Insight}-HMXT increased from 1.55 to 3.65~hr.

\subsection{Spectral analysis of the persistent emission} \label{subsec:3.2}

We fitted the simultaneous \hxmt\ LE, ME, and HE spectra using \textsc{xspec} version 12.12.1 \citep{1996ASPC..101...17A}. The \hxmt\ spectra were grouped by a factor of 5 using the tool {\tt grppha}. All uncertainties of the spectral parameters are provided at a $1\sigma$ confidence level for a single parameter. 

We adopted the energy ranges of 2--10~keV, 8--20~keV, and 30--100~keV for \hxmt\ LE, ME, and HE spectra, respectively. We fit all the spectra by using the thermally Comptonized continuum, {\tt nthcomp}, modified by the interstellar absorption described by the model, {\tt tbabs}. The full model is {\tt tbabs$\times$nthcomp} in \textsc{xspec}. The blackbody seed photons were assumed for \texttt{nthcomp}.

To account for cross-calibration uncertainties between different instruments, a multiplication factor, {\tt constant} in \textsc{xspec}, is included in the fits. The factor is fixed at unity for \hxmt/LE and freed for \hxmt/ME/HE. The other free parameters are: the asymptotic power-law photon index, $\Gamma$, the electron temperature, $kT_{\rm e}$, the temperature of the soft seed photons, $kT_{\rm bb}$, assuming the blackbody type of seed photons from NS, the normalization, and the hydrogen column density, $N_{\rm H}$. These parameters of ME and HE were tied with the same values for LE. 

\begin{figure*}
\centerline{\includegraphics[width=1\linewidth]{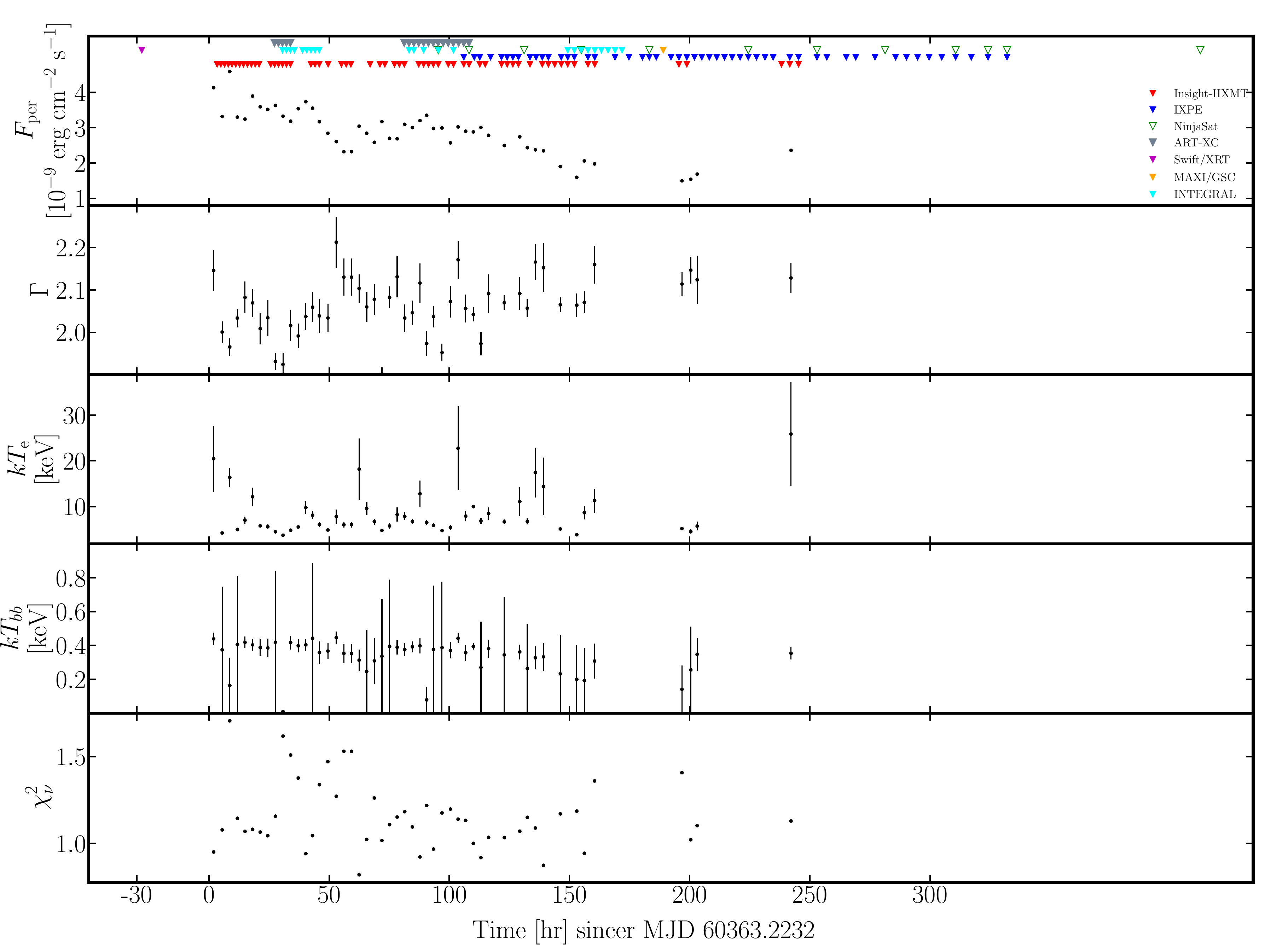}}
    \caption{Evolution of persistent spectrum with \hxmt\,data with the bolometric flux calculated in 0.5--250 keV. The black points represent the parameters of model \texttt{constant$\times$TBabs$\times$nthcomp} with the fixed centroid. The onset time of all type I bursts is marked as a triangle in the top panel.}
    \label{fig:3}
\end{figure*}

The fitted parameters throughout the outburst are shown in Figure~\ref{fig:3}. All persistent spectra can be fitted with $\chi^{2}_{\nu}\sim0.82\text{--}1.70$, with a degrees-of-freedom range of 176.0--288.0. The best-fitted results of a few spectra have $\chi^{2}_{\nu}>1.5$, which were from the fluctuation of \hxmt/HE spectra. We calculated the unabsorbed bolometric flux in the 0.5--250~keV range by using the tool \texttt{cflux}. The temperature of seed photons was around 0.4~keV. 
The hydrogen column density did not change much, and the mean value is $(1.87\pm0.18)\times10^{22}~{\rm cm^{-2}}$; thus it was fixed during the fitting. \citet{2024ApJ...968L...7N} and \citet{2024arXiv240800608P} reported a higher $N_{\rm H}$ of $\sim2.9\times10^{22}~{\rm cm^{-2}}$, which may be attributed to the broader energy coverage of the \hxmt\ spectra. The asymptotic power-law photon index was in the range of 1.9--2.2, which explains the visibility of hard X-ray emissions during the whole outburst. The electron temperature was in the range 10--25~keV. The bolometric flux decreased from a peak value of $4.59\times10^{-9}~{\rm erg~cm^{-2}~s^{-1}}$ to $2.36\times10^{-9}~{\rm erg~cm^{-2}~s^{-1}}$ at MJD~60373.3.

\subsection{Time-resolved burst spectral analysis} \label{subsec:3.3}
We performed time-resolved spectral fits on each burst observed with \hxmt. Bursts \# 35 and \# 45 occurred during a period of unusually high background levels; therefore, their burst spectra were polluted and ignored in the burst spectral analysis. So, we analyzed 58 type I X-ray bursts from the \hxmt\ observations. For ME bursts, we extracted the spectra with an exposure time of 1--4~s to ensure at least 100 counts in each spectrum. We used the same time intervals to extract LE burst spectra if the LE data were available. The burst spectra were grouped using ftool \texttt{grappha} with a minimum of 10 counts per channel.

To fit the burst spectra, we used an absorbed blackbody model, \texttt{TBabs$\times$bbodyrad}. We regarded the pre-burst spectra as background, which was assumed to be invariant during the bursts. The spectral parameters include the blackbody temperature $kT_{\rm bb}$ and normalization ($R^2_{\rm km}/D^2_{10}$),
where $R_{\rm km}$ is the apparent burst radius in km, and $D_{10}$ is the distance to the source in units of 10~kpc. 
For each burst, we fixed the hydrogen column density at $1.87\times10^{22}~{\rm cm^{-2}}$. For the joint LE and ME spectral fitting, to account for the instrumental calibration, a constant was added to the model, which was fixed at 1 for LE, and set free for ME. The model fitted most of the burst spectra well with $\chi^2_{\nu}\sim0.5\text{--}1.5$.

\begin{figure}
  \centerline{\includegraphics[width=1\linewidth]{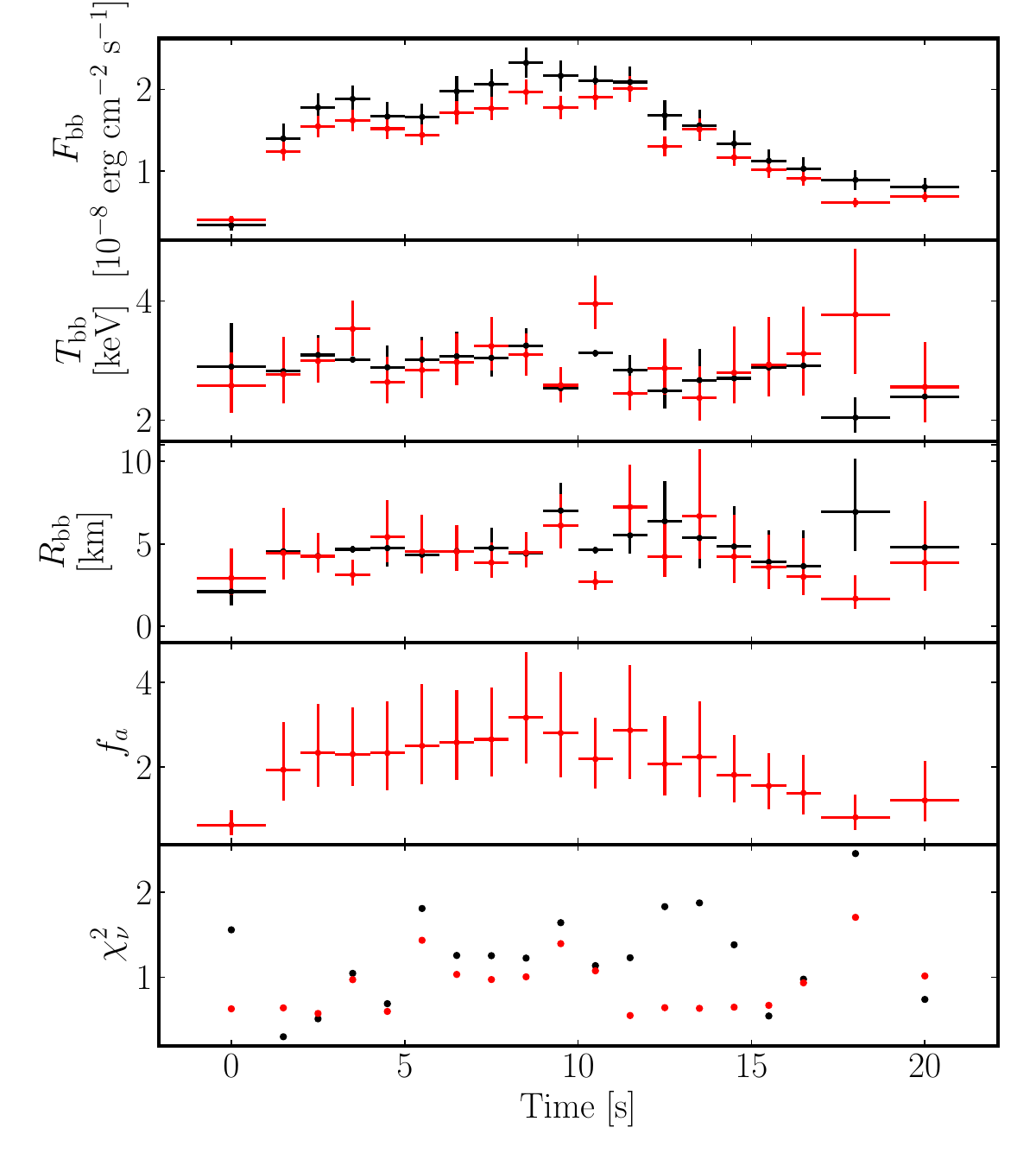}}
    \caption{Time-resolved spectroscopy using the 
    variable persistent flux 
    model (red dot) and the blackbody model (black dot) for joint LE and ME spectra of burst $\#$10.
    The panels are, from top to bottom, the bolometric burst flux $F_{\rm bb}$; blackbody temperature $T_{\rm bb}$; blackbody radius $R_{\rm bb}$; multiplicative term for the persistent flux $f_a$ (only relevant for the variable persistent flux method); and fit statistic, $\chi^2_\nu$.}
    \label{fig:4}
\end{figure}

 Please note that the assumption of invariant persistent emission during X-ray bursts may not hold. Bursts can exhibit higher blackbody temperatures compared to nonburst spectra, potentially producing different Comptonized emissions. Consequently, parameters such as $kT_{\rm bb}$ and the normalization of \texttt{nthcomp} may vary. For some burst spectra with large $\chi^2_{\nu}$, for simplicity, we 
used instead the variable persistent flux 
method to improve the fitting. We assumed that only the normalization of the persistent component can change during bursts, and the spectral shape remains constant. Then only the instrumental background was subtracted. The model is \texttt{constant$\times$TBabs$\times$(bbodyrad+$f_a\times$nthcomp)}. The \texttt{nthcomp} accounts for the persistent emission during the burst, and the parameters were fixed to the best-fitted values for each observation, listed in Table \ref{Table:2}. Parameter $f_a$ is a free scaling factor that is used to account for the persistent emission variation during the burst. $f_a = 1$ means that the normalization of the persistent emission during the burst is unchanged compared to the value before the burst. By fitting all the burst spectra, we found that the 
variable persistent flux
model can improve the fit, but most of the parameters are consistent with the blackbody model at a $1\sigma$ confidence level. See, e.g. Figure~\ref{fig:4} for burst \#10, which has the most significant improvement. In this work, we only reported the results from the blackbody model.

\begin{figure*}
    \centering
    \includegraphics[width=1\linewidth]{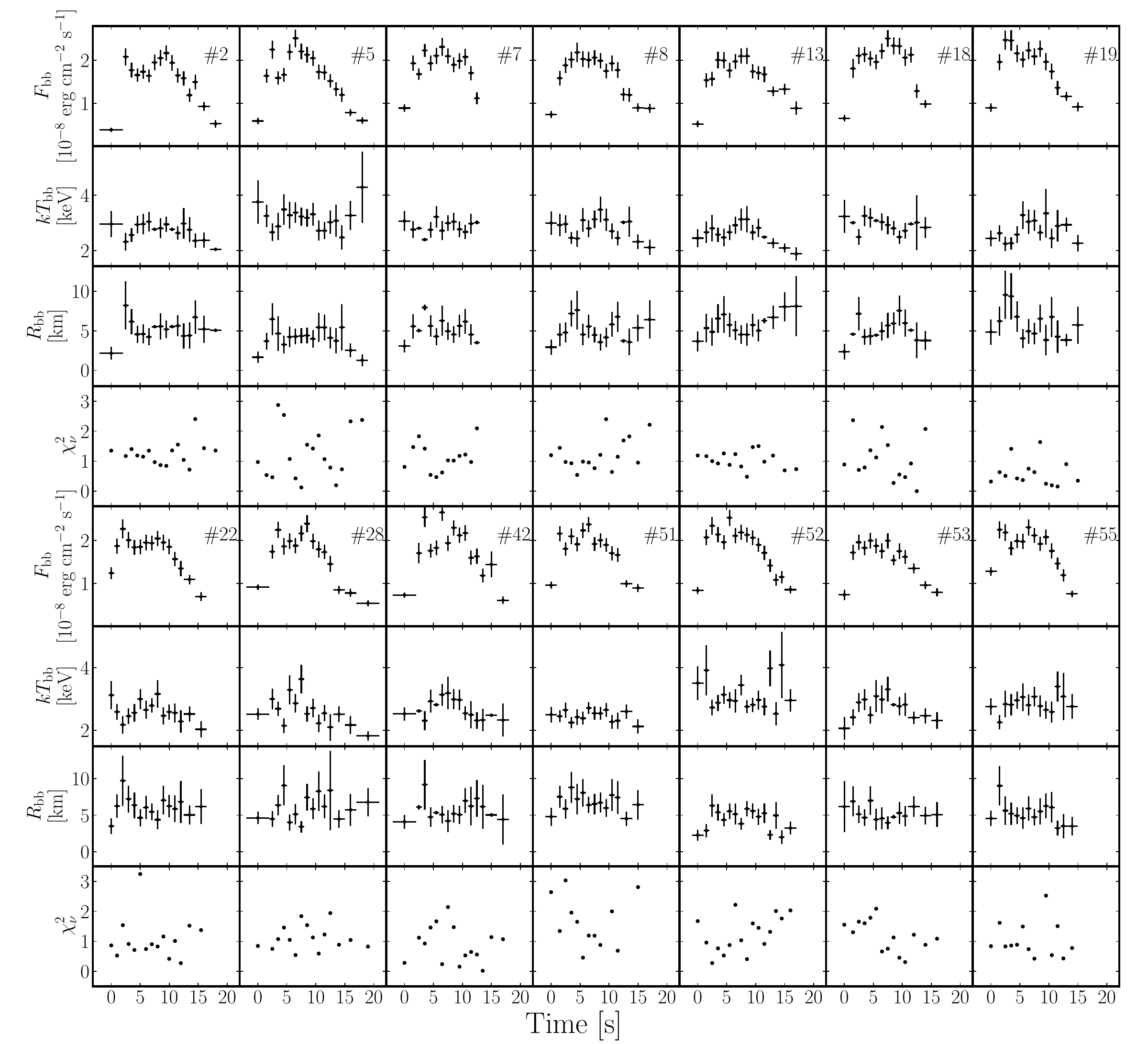}
    \caption{Time-resolved spectroscopy using the \texttt{TBabs$\times$bbodyrad} for PRE bursts \#2, \#5, \#7, \#8, \#13, \#18, \#19, \#22, \#28, \#42, \#51, \#52, \#53, and \#55. For each panel, from
top to bottom, we exhibit the burst bolometric flux, $F_{\rm bb}$; blackbody temperature, $kT_{\rm bb}$, blackbody radii, $R_{\rm bb}$, which were calculated
using a distance of 10.03~kpc, and goodness of fit per degree of freedom, $\chi^{2}_{\nu}$, respectively.}
\label{fig:PRE}
\end{figure*}

Based on the time-resolved spectra, we found that 14 bursts, including bursts \#2, \#5, \#7, \#8, \#13, \#18, \#19, \#22, \#28, \#42, \#51, \#52, \#53, and \#55 exhibit PRE (see Figure~\ref{fig:PRE}). For these 14 bursts, the blackbody radius increased at the burst start and reached a maximum of around 10~km (for $d=10.03$~kpc, see Sect. \ref{subsec:4.2}). When the photosphere fell back to the NS surface, the color temperature increased to its peak at $\sim3$~keV and the apparent radius decreased to its local minimum, which corresponds to the ``touchdown'' moment. Afterward, the color temperature decreased, and the apparent radius was found to be consistently maintained at a constant level of $6.22\pm1.53$~km, which is an average value derived from all the PRE bursts.

All bursts were characteristic of blackbody flux with fast rising and exponential decaying, and spectral softening after the burst peak. We calculated the burst fluence, $f_{\rm b}$, from the sum of the product of blackbody flux and its exposure time.
From the parameters fitted above, the burst decay timescale is obtained by $\tau=f_{\rm b}/F_{\rm peak}$, where $F_{\rm peak}$ is the peak flux. All burst parameters are listed in Table~\ref{Table:2}.

\section{Discussion} 
\label{sec:4}

In this paper, we analyzed the broadband \hxmt\ observations of the newly discovered AMXP \psrtar\ during its 2024 outburst. We found 60 type I (thermonuclear) X-ray bursts from \hxmt/ME, and 37 of them were also observed by \hxmt/LE. The persistent spectra in the energy range of 2--150 keV can be well fitted with the \texttt{tbabs$\times$nthcomp} model. Most of the bursts showed a similar profile, a fast rise in 2--3~s followed by a plateau lasting 7--10~s and then an exponential decay.

We performed a detailed time-resolved spectral analysis of 58 type I X-ray bursts. We found that the 
variable persistent flux
model did not improve the residuals of the \texttt{bbodyrad} model significantly, which may be caused by the limited signal-to-noise ratio of the spectra. By stacking the light curves, as reported in Sect.~\ref{subsec:4.1} we measure a deficit in the HE light curve. From the time-resolved spectra, 14 bursts show PRE, and the burst composition and source distance are determined in Sect.~\ref{subsec:4.2}. In Sect.~\ref{subsec:4.3}, we discuss the relation between the burst recurrence time and the local mass accretion rate. We compare the calculated recurrence time with the actual recurrence time, which further reconfirms the fuel composition of hydrogen and helium.

\subsection{The stacking of burst light curves} \label{subsec:4.1}

\begin{figure}
    \centerline{\includegraphics[width=1\linewidth]{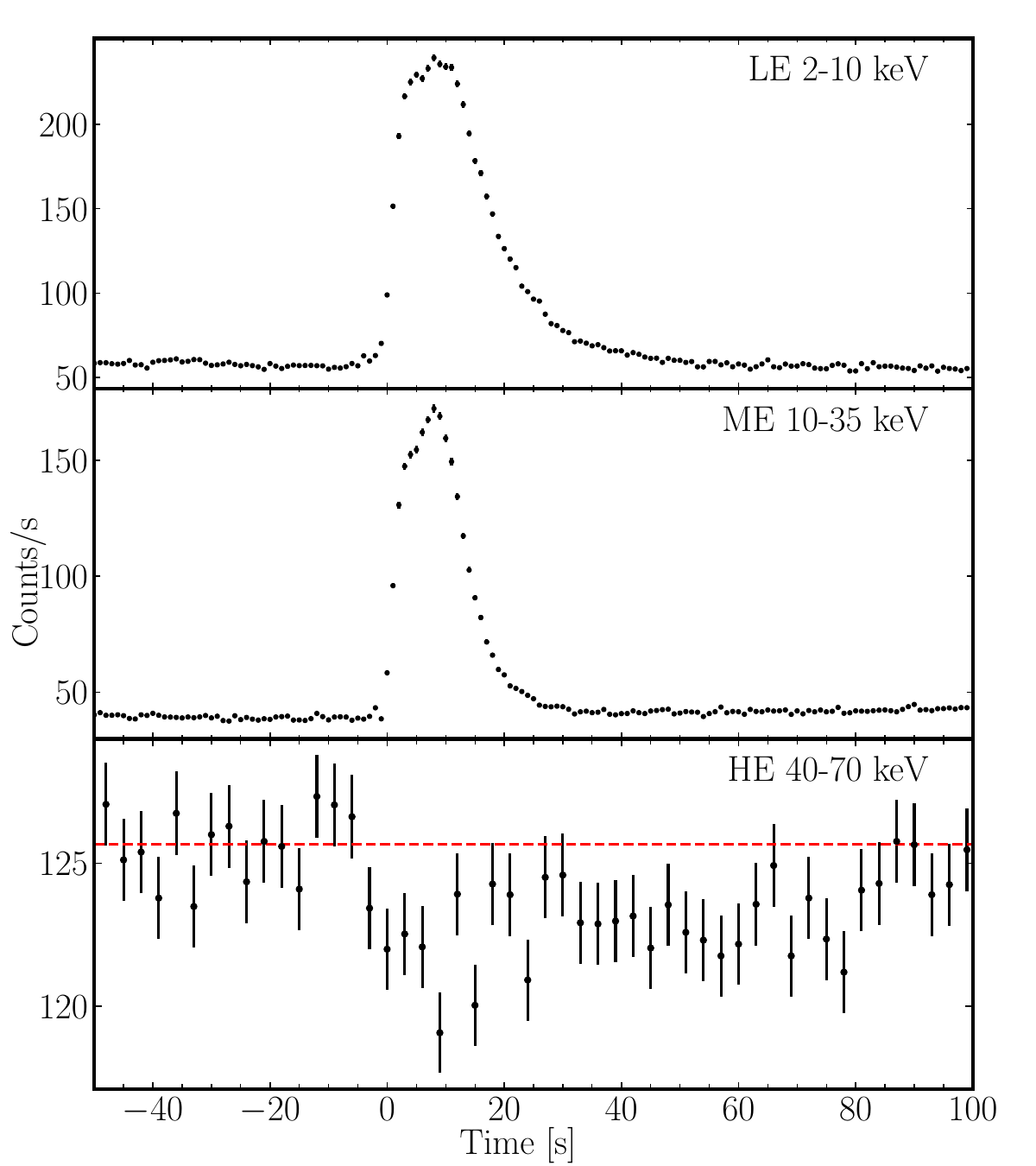}}
    \caption{From top to bottom panels, we show the LE, ME, and HE light curves during bursts in 2--10 keV, 10--35 keV, and 40--70 keV, respectively. The bin size is 1~s for LE and ME, and 3~s for HE.}
    \label{fig:5}
\end{figure}

The spectra of LMXBs, including both black hole and NS X-ray binaries (XRBs), are generally composed of soft/thermal (e.g., blackbody or disk blackbody) and hard/Comptonizated components. The hard X-rays in XRBs are produced via inverse Compton scattering of the soft photons off the hot electrons in the corona. The burst spectrum is usually modeled by a blackbody model of characteristic temperature in 0.5--3~keV. The burst emission can reach $L_{\rm Edd}$ and dominates the total emission at energies well below $\sim30$~keV, above which the persistent emission from the purported corona dominates. Therefore, the \hxmt/HE observations in the 40--70~keV energy band allow us to study the possible effects of the bursts on the corona. 

The LE and ME burst light curves have a similar profile, which can be stacked to probe the effects of burst-induced corona cooling with improved statistics. We combined the LE, ME, and HE light curves for all the bursts in the 2--10, 10--35, and 40--70~keV energy ranges, respectively, between -50 and 100~s relative to the burst onset time \cite[following][]{2012ApJ...752L..34C}. 

As shown in Figure~\ref{fig:5}, the HE flux dropped at the same time that LE and ME detected an increase in burst flux. As the burst decays, the HE flux returns to pre-burst levels. The mean HE pre-burst rate was $\sim126~\rm cts~s^{-1}$, including a background contribution of $\sim 121\rm~cts~s^{-1}$ and a persistent (source) contribution of $\sim5\rm~cts~s^{-1}$. At the peak of the burst, the HE rate decrement reaches a maximum of $\sim6\pm1.5\rm~cts~s^{-1}$ corresponding to a significance of $4\sigma$, which was $\sim120\%\pm30\%$ of the total persistent flux in 40--70~keV. \cite{2022ApJ...936L..21C} and \cite{2024ApJ...966L...3J} proposed that the deficit fraction might be systematically lower for pulsating compared to nonpulsating LMXBs. However, for \psrtar, the reduced persistent emission in its hard energy band cannot give a reasonable deficit fraction.

A cross-correlation analysis was carried out between the ME light curve in 10--35~keV and the HE light curve in 40--70~keV with a bin size of 1 s. The HE light curve was anticorrelated with the ME light curve, with the minimum value at $0.81\pm0.58$~s, which indicates the hard X-ray deficit lagged behind the burst emission.

Similar phenomena have also been reported in the type I X-ray bursts of IGR J17473--2721 \citep{2012ApJ...752L..34C}, Aql X--1 \citep{2013ApJ...777L...9C}, 4U 1636--536 \citep{2018ApJ...864L..30C}, and MAXI J1816--195 \citep{2022ApJ...936L..21C}. The deficit in hard X-rays during bursts is explained as the cooling of the corona by the burst, which provides an intense shower of soft X-rays to cool the hot corona via the Compntonization process. When bursts do not occur, the corona cooling is driven by the soft photons from the disk or boundary layer. When bursts occur, their soft photons overwhelm those from the disk and provide additional cooling in a typical timescale of tens of seconds. The anti-correlation between hard and soft X-rays suggests that the corona can be cooled down and recover quite fast within a few seconds. Such a short timescale for the corona recovery is inconsistent with the disk evaporation model in which the formation of a corona is driven and energized by the disk accretion with a typical timescale of days. As previously discussed by \citet{2000Sci...287.1239Z} and \citet{2007HiA....14...41Z}, an alternative magnetic field reconnection model can explain such rapid evolution. Magnetic field reconnection in the inner disk region can release the kinetic energies in the rotating disk to heat the corona with the Keplerian orbital timescale of milliseconds, comparable to the case of \psrtar.

\subsection{The source distance and X-Ray burst fuel} \label{subsec:4.2}

From the persistent and burst spectral results, we can determine the burst fuel and source distance. We calculate $\alpha$ factor to verify the burst fuel composition, 
\begin{equation}
\alpha = \frac{\Delta T_{\rm rec}F_{\rm per}}{f_{\rm b}},
\end{equation}
which is the ratio between the persistent to the burst fluence. From the average values of burst fluence, persistent flux, and $\Delta T_{\rm rec}$, we found the average $\alpha = 71 \pm 7$, which is consistent with the mixed hydrogen and helium fuel. We can also calculated the $\alpha$ as
\begin{equation}
\alpha = \frac{Q_{\rm grav}}{Q_{\rm nuc}}\frac{\xi_{\rm b}}{\xi_{\rm p}}(1+z),
\end{equation}\label{Eq:alpha}
where $Q_{\rm grav}=c^{2}z(1+z)\approx GM_{\rm NS}/R_{\rm NS}$, $\xi_{\rm b}$ and $\xi_{\rm p}$ account for the anisotropy of burst and persistent emission, respectively. The gravitational redshift at the photosphere $1+z = (1-2GM_{\rm NS}/Rc^{2})^{-1/2}= 1.259$ for $M_{\rm NS} = 1.4M_{\odot}$, and $R = R_{\rm NS} = 11.2$ km. For a fuel layer consisting of mixed hydrogen and helium, $Q_{\rm nuc}$ depends on the mean hydrogen fraction, $\overline{X}$, at ignition, which is given by \citep{2019ApJ...870...64G}
\begin{align}
\label{Eq:3}
Q_{\rm nuc} & \approx 1.31 + 6.95\overline{X} - 1.92\overline{X}^{2} \; {\rm MeV} \; {\rm nucleon^{-1}}.
 \end{align}
By substituting the linear expression in Equation~(\ref{Eq:3}) for simplicity, it can be estimated \citep{Galloway22}
\begin{align}
    \overline{X} &=z\frac{155}{\alpha}\frac{\xi_{\rm b}}{\xi_{\rm p}} - 0.223.
\end{align}
We assumed that the anisotropy factors $\xi_{\rm b}$ and $\xi_{\rm p}$ are both unity. So we obtained the mean $\overline{X}$ = 0.342 $\pm$ 0.033.

If we assume the observed peak flux of PRE bursts corresponding to the Eddington limit, the distance to the source can be measured as,
\begin{align}
d & =\left(
\frac{L_{\rm Edd, \;\infty}}{4\pi\xi_{b}F_{\rm pk,\;RE}} \right) ^{1/2} \notag \\
& = 8.83\left( \frac{\xi_{b}F_{\rm pk,\; RE}}{3\times10^{-8} {\rm \; erg \; cm^{-2} \; s^{-1}}} \right)^{-1/2}\left(\frac{M_{\rm NS}}{1.4M_{\odot}}\right)^{1/2}  \notag\\
& \times \left(\frac{1+z}{1.259}\right)^{-1/2}\left(1+X\right)^{-1/2}{\rm kpc},
\end{align}
where $F_{\rm pk,\;PRE}$ is the mean peak flux ($2.33\pm0.17)\times \rm 10^{-8}~erg~cm^{-2}~s^{-1}$ of 14 PRE bursts, $X$ is the mass fraction of hydrogen in the atmosphere. The effective redshift measured at the peak of a PRE burst may be lower than the value at the NS surface, while the photosphere is expanded during the radius expansion episode (i.e., $R \geqslant R_{\rm NS}$). Evidence suggests that even if the atmosphere contains hydrogen, the hydrogen-rich material is typically blown off during the PRE phase, exposing the underlying helium layer \citep[see, e.g., ][]{2006ApJ...639.1033G,2019ApJ...885L...1B,2023HEAD...2011611G}. Then we obtained the distance $d = 10.03\pm0.71$~kpc for $X = 0$. Hereafter, we use 10.03 kpc as the distance to the source.

\begin{figure}
    \centerline{\includegraphics[width=1.2\linewidth]{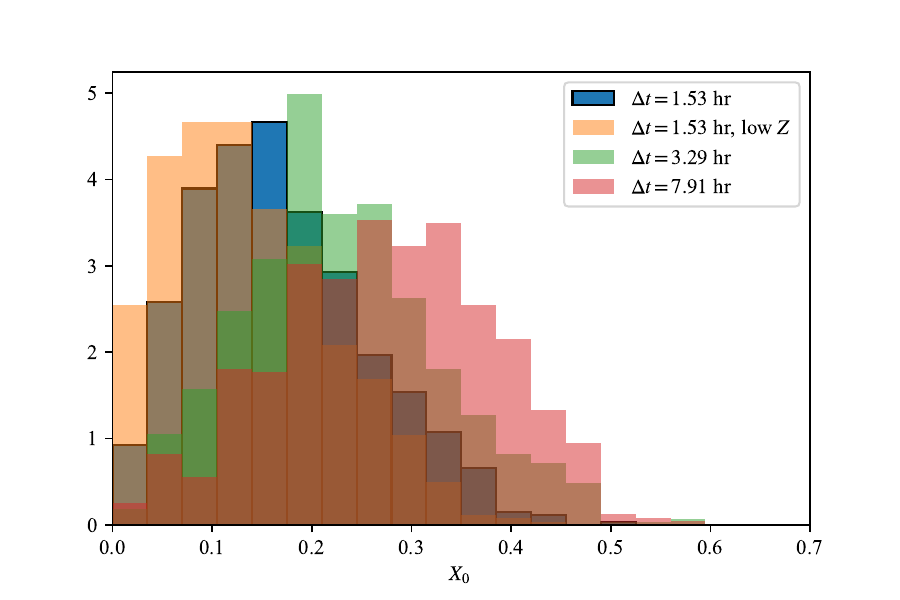}}
    \caption{Inferred probability density function of the H-fraction $X_0$ for the burst fuel, based on three pairs of bursts with different recurrence times. The shortest pair of bursts provides the most stringent constraints on $X_0$, strongly suggesting the burst fuel is H-deficient. For the same pair, we also plot the inferred $X_0$ for low CNO metallicity, $Z_{\rm CNO}=0.005$ rather than the roughly solar value of 0.02.}
    \label{fig:xdist}
\end{figure}

We selected three pairs of bursts for further analysis; two were observed with {\it Insight}-HXMT (\#4 \& 57 in Table~\ref{Table:2}), and an additional pair was observed by {\it IXPE}, with recurrence times 1.53, 3.29, and 7.91~hr, respectively. For each pair, we calculated the $\alpha$ value (equation \ref{Eq:alpha}), based on the estimated persistent flux at the time of the burst, and the burst fluence. We then inferred the average H-fraction in the burst fuel via estimates of $Q_{\rm nuc}$ and $\overline{X}$, for the average quantities above. We also incorporated the effects of the persistent and burst flux anisotropy, using the {\sc concord} suite of tools following \cite{Galloway22}. We adopted the inclination inferred by \cite{2024arXiv240800608P} from the phase-resolved Stokes parameters, of $i = (74.1^{+5.8}_{-6.3})^\circ$. For a roughly solar metallicity ($Z_{\rm CNO}=0.02$) the fuel H-fraction, $X_0$, is significantly below solar (Fig. \ref{fig:xdist}). In fact, for the most constraining pair of bursts (with $\Delta t=1.53$) $X_0\lesssim0.4$. Reducing $Z_{\rm CNO}$ shifts this distribution to even lower values of $X_0$, so we consider the upper limit robust unless the CNO metallicity is instead substantially {\it super}-solar.

\subsection{The burst recurrence time} \label{subsec:4.3}

\begin{figure}
    \centering
    \includegraphics[width=1\linewidth]{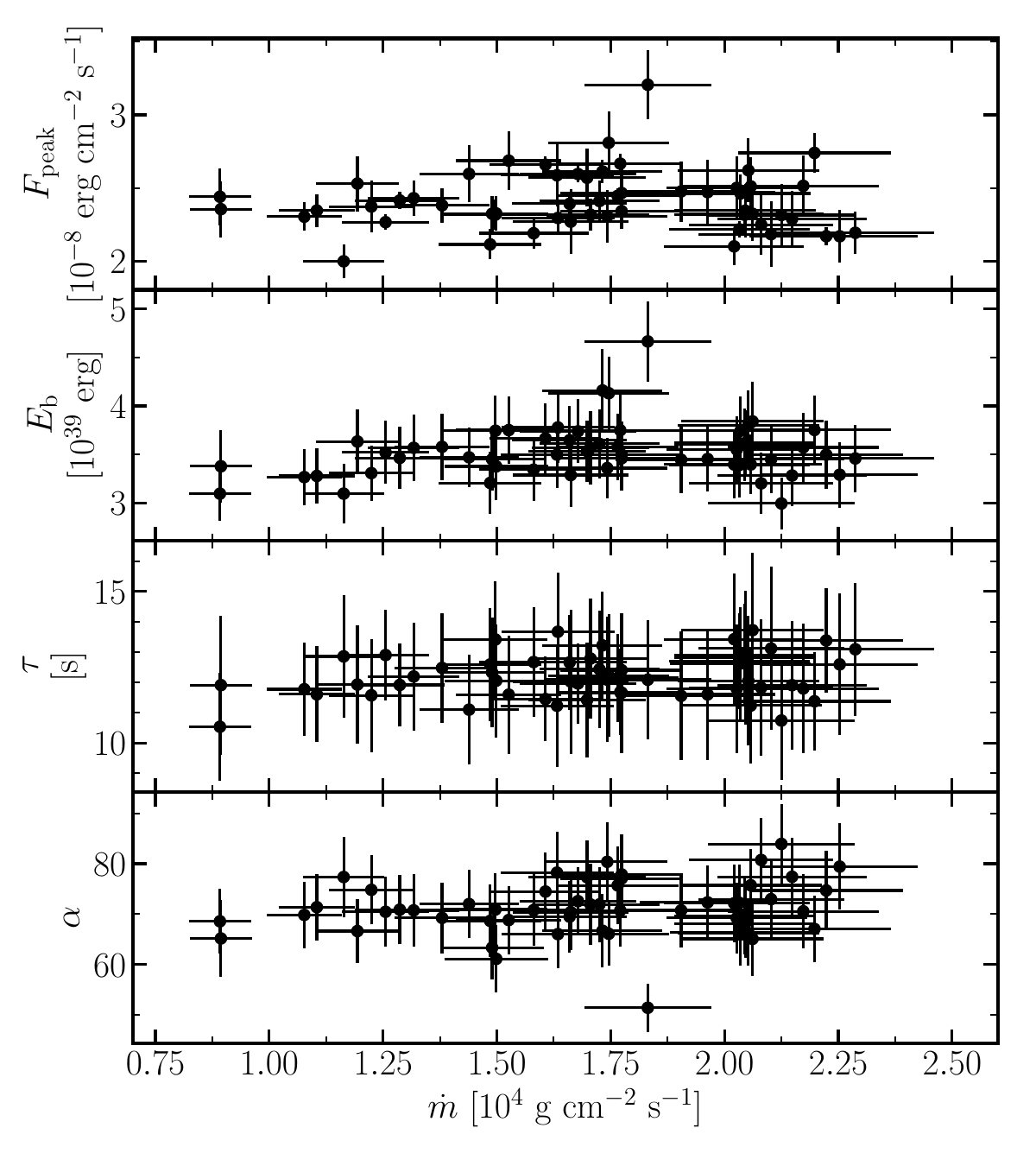}
    \caption{Burst parameters, peak flux, total released energy, and $\alpha$, from top to bottom panel, related to the local mass accretion rate. The $\alpha$ value of the last three bursts is not shown due to the inaccuracy in the pre-burst persistent flux. }
    \label{fig:mdot_fpeak}
\end{figure}

\begin{figure}
    \centering
    \includegraphics[width=1.1\linewidth]{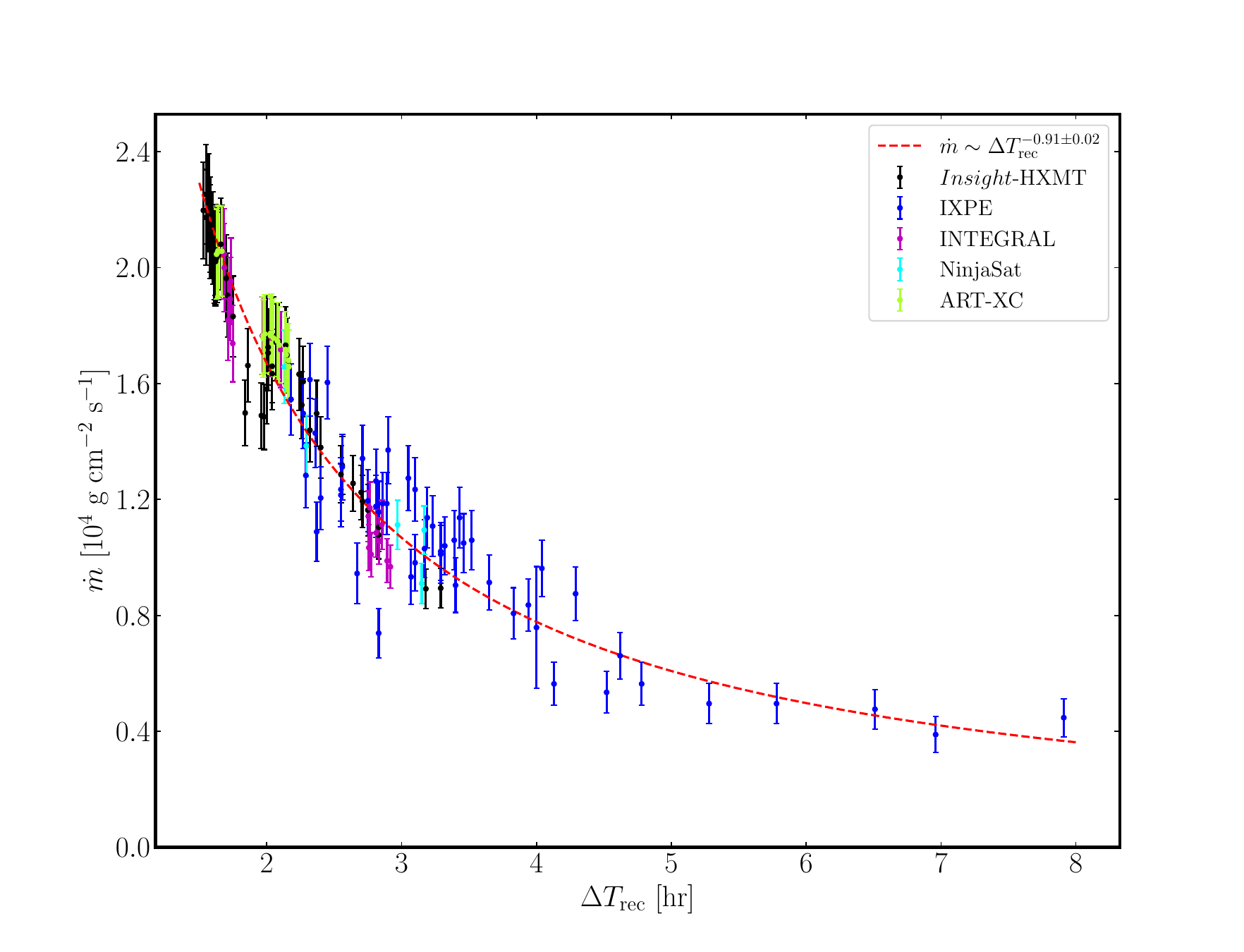}
    \caption{Relation between the recurrence time and local mass accretion rate. The red dashed line represents the best-fitted power law $\dot{m}\sim \Delta T_{\rm rec}^{-0.91\pm0.02}$.}
    \label{fig:7}
\end{figure}

We found that the burst recurrence time increased from 1.55 to 3.65 hr as the persistent flux decreased. In the following, we verifiy the relation between recurrence time and local accretion rate.
The local accretion rate is determined via \citep{2008ApJS..179..360G},
\begin{align}
\label{Eq:6}
    \dot{m} & =\frac{L_{\rm per}[1+z]}{4\pi R^{2}_{\rm NS}(GM_{\rm NS}/R_{\rm NS})}\notag \\
    & \approx 6.7 \times 10^{3} \left(\frac{F_{\rm per}}{10^{-9} \; {\rm erg} \; {\rm cm^{-2}} \; {\rm s^{-1}}}\right)\left(\frac{d}{10 \; {\rm kpc}}\right)^{2} \notag\\
    & \times \left(\frac{M_{\rm NS}}{1.4M_{\odot}}\right)^{-1}\left(\frac{1+z}{1.259}\right)\left(\frac{R_{\rm NS}}{\rm 11.2 \;km}\right)^{-1} {\rm g} \;{\rm cm^{-2}} \;{\rm s^{-1}},
\end{align} 
where $F_{\rm per}$ is the persistent flux. In Fig.~\ref{fig:mdot_fpeak}, we show the evolution of peak flux, the total energy released, the decay time, and $\alpha$ as a function of $\dot{m}$. We found that all these parameters were almost unchanged from the peak to the end of the outburst, the same as shown from burst light curves. We estimated the ignition depth at the onset of the burst with the equation
\begin{align}
\label{Eq:7}
y_{\rm ign} & =  \frac{4\pi f_{\rm b} d^{2} (1+z)}{4\pi R_{\rm NS}^{2} Q_{\rm nuc}},
\end{align}
which are listed in Table \ref{Table:2}. 

Since the complete burst sample was obtained without GTI filtering, the pre-burst persistent emission flux for some bursts could not be obtained by the standard spectral fitting process. By applying the UnivariateSpline method, we interpolated the persistent flux from Sect.~\ref{subsec:3.2} to obtain the pre-burst persistent emission. Then, the local mass accretion rates for bursts from \hxmt, {\it IXPE}, and \Integ, are calculated. Since the uncertainties of recurrence time are much smaller than the measured value, we only considered the uncertainties of the local mass accretion rate. The observed recurrence time increased with decreasing of $\dot{m}$ as $\dot{m}\sim \Delta T_{\rm rec}^{-1.09\pm0.03}$, or inversely $\Delta T_{\rm rec}\sim \dot{m}^{-0.91\pm0.02}$; see Figure~\ref{fig:7}. By assuming that the bursts occurred under the same physical conditions, the accreted matter between the two bursts should be the same. Therefore, it is expected that the local accretion rate, $\dot{m}$, is related to the burst recurrence time by $\Delta T_{\rm rec}\sim\dot{m}^{-1}$ \citep{2018A&A...620A.114L}. The power-law index of $-0.91$ in \psrtar\ measured by \hxmt\ slightly deviates from the expectation. We also note that the X-ray bursts in \psrtar\ detected by {\it IXPE} and NinjaSat showed similar trends of $\Delta T_{\rm rec}\sim C^{-0.8}$ and $\Delta T_{\rm rec}\sim C^{-0.84}$, respectively, although the direct measurement of the local mass accretion rate was difficult \citep{2024arXiv240800608P, takeda2024ninjasatmonitoringtypeixray}. As simulated by \citet{Dohi24}, the power-law index depends on the NS mass, radius, and the equation of state. A smaller power-law index ($<1$) indicates NS mass higher than $2M_\odot$ (see also the discussion in \citet{takeda2024ninjasatmonitoringtypeixray}). The $\dot{m}-\Delta t$ relation can be used to constrain the NS equation of state by extending the simulation to the $M>2M_\odot$ regime to both satisfy the observed power-law index and the maximum mass of the NS equation of state.

\section{Summary} \label{sec:5}
We detected 60 type I X-ray bursts from SRGA J144459.2--604207 with \hxmt\ observations. The persistent spectra of all bursts were well fitted by \texttt{TBabs$\times$nthcomp}, showing a similar spectral shape. A shortage of hard X-rays in the 40--70 keV range was detected by stacking the \hxmt\ light curves of 37 type I X-ray bursts with similar profiles and count rates. We analyzed the time-resolved spectra of all bursts and found that 14 of 60 type I X-ray bursts are PRE bursts. We obtained a distance of $10.03\pm0.71$~kpc for this source. Based on the depth of ignition, the local accretion rate, and the recurrence time, we proposed that these X-ray bursts were ignited in a mixture of hydrogen and helium environments.

\begin{acknowledgments}
We thank the referee for the valuable comments, which improved our manuscript.
We appreciate Celia Sanchez-Fernandez and Sergey Molkov for sharing the onset time of INTEGRAL and SRG/ART-AC bursts, respectively. This work was supported by the Major Science and Technology Program of Xinjiang Uygur Autonomous Region (No. 2022A03013-3). Z.L. and Y.Y.P. were supported by National Natural Science Foundation of China (12103042, 12273030, U1938107). This work made use of data from the \hxmt\ mission, a project funded by China National Space Administration (CNSA) and the Chinese Academy of Sciences (CAS).

\end{acknowledgments}

\clearpage
\startlongtable
\centerwidetable
\begin{deluxetable*}{ccccccccccc}

\tabletypesize{\footnotesize}

\tablecaption{Table 1}

\tablenum{1}
\label{Table:2}
\tablehead{\colhead{NO.} & \colhead{Obs id$^{a}$} & \colhead{Burst Onset} & \colhead{$F_{\rm peak}$} & \colhead{$f_{\rm b}$} & \colhead{PRE} & \colhead{$\Delta T^{\rm b}_{\rm rec}$} & \colhead{$\tau$} & \colhead{$\alpha$} & \colhead{$\dot{m}$} & \colhead{$y_{\rm ign}$} \\ 
\colhead{(\#)} & \colhead{} & \colhead{(MJD)} & \colhead{($10^{-8}~{\rm erg}~{\rm s}^{-1}~{\rm cm}^{-2}$)} & \colhead{($10^{-7}~\rm{erg~cm^{-2}}$)} & \colhead{} & \colhead{(hr)} & \colhead{(s)} & \colhead{} & \colhead{($10^4~\rm{g~cm^{-2}~s^{-1}}$)} & \colhead{($10^8~\rm{g~cm^{-2}}$)} } 

\startdata
1 & 101 & 60363.36440 & 2.19$\pm$0.15 & 2.87$\pm$0.29 & N & $\--$ & 13.09$\pm$2.19 & $\--$ & 2.29$\pm$0.17 & 0.83$\pm$0.14 \\
2 & 102 & 60363.42902 & 2.17$\pm$0.18 & 2.73$\pm$0.28 & Y & 1.55 & 12.59$\pm$2.34 & 77.42$\pm$8.33 & 2.25$\pm$0.17 & 0.79$\pm$0.14 \\
3 & 102 & 60363.49440 & 2.17$\pm$0.06 & 2.90$\pm$0.29 & N & 1.57 & 13.38$\pm$1.72 & 71.57$\pm$7.54 & 2.22$\pm$0.17 & 0.84$\pm$0.14 \\
4 & 103 & 60363.55817 & 2.74$\pm$0.13 & 3.12$\pm$0.29 & N & 1.53 & 11.37$\pm$1.63 & 63.54$\pm$6.31 & 2.20$\pm$0.17 & 0.90$\pm$0.15 \\
5 & 103 & 60363.62261 & 2.52$\pm$0.21 & 2.97$\pm$0.29 & Y & 1.55 & 11.80$\pm$2.14 & 65.71$\pm$6.85 & 2.17$\pm$0.17 & 0.85$\pm$0.15 \\
6 & 104 & 60363.68849 & 2.29$\pm$0.19 & 2.73$\pm$0.26 & N & 1.58 & 11.90$\pm$2.12 & 70.18$\pm$7.04 & 2.15$\pm$0.16 & 0.79$\pm$0.13 \\
7 & 104 & 60363.75415 & 2.32$\pm$0.21 & 2.49$\pm$0.22 & Y & 1.58 & 10.74$\pm$1.95 & 73.90$\pm$7.02 & 2.12$\pm$0.16 & 0.72$\pm$0.12 \\
8 & 105 & 60363.82095 & 2.19$\pm$0.22 & 2.87$\pm$0.29 & Y & 1.60 & 13.11$\pm$2.69 & 63.67$\pm$6.83 & 2.10$\pm$0.16 & 0.83$\pm$0.14 \\
9 & 105 & 60363.89030 & 2.25$\pm$0.21 & 2.66$\pm$0.26 & N & 1.66 & 11.82$\pm$2.25 & 75.14$\pm$7.80 & 2.08$\pm$0.16 & 0.77$\pm$0.13 \\
10 & 106 & 60363.95762 & 2.32$\pm$0.19 & 3.19$\pm$0.34 & N & 1.62 & 13.72$\pm$2.57 & 67.51$\pm$7.57 & 2.06$\pm$0.16 & 0.92$\pm$0.16 \\
11 & 106 & 60364.02469 & 2.35$\pm$0.18 & 3.00$\pm$0.28 & N & 1.61 & 12.81$\pm$2.21 & 73.62$\pm$7.29 & 2.05$\pm$0.16 & 0.87$\pm$0.14 \\
12 & 107 & 60364.09186 & 2.22$\pm$0.06 & 2.82$\pm$0.28 & N & 1.61 & 12.69$\pm$1.59 & 74.53$\pm$7.77 & 2.03$\pm$0.15 & 0.81$\pm$0.14 \\
13 & 108 & 60364.29415 & 2.10$\pm$0.13 & 2.82$\pm$0.29 & Y & 4.85/1.62 & 13.42$\pm$2.18 & 72.98$\pm$7.84 & 2.02$\pm$0.15 & 0.81$\pm$0.14 \\
14 & 109 & 60364.36159 & 2.50$\pm$0.21 & 2.95$\pm$0.28 & N & 1.62 & 11.79$\pm$2.13 & 71.74$\pm$7.23 & 2.03$\pm$0.15 & 0.85$\pm$0.14 \\
15 & 109 & 60364.42943 & 2.46$\pm$0.13 & 3.11$\pm$0.29 & N & 1.63 & 12.62$\pm$1.86 & 69.18$\pm$6.88 & 2.04$\pm$0.15 & 0.90$\pm$0.15 \\
16 & 110 & 60364.49771 & 2.32$\pm$0.06 & 2.99$\pm$0.31 & N & 1.64 & 12.90$\pm$1.68 & 70.81$\pm$7.76 & 2.04$\pm$0.16 & 0.86$\pm$0.15 \\
17 & 110 & 60364.56604 & 2.62$\pm$0.22 & 3.16$\pm$0.29 & N & 1.64 & 12.05$\pm$2.13 & 63.09$\pm$6.20 & 2.05$\pm$0.16 & 0.91$\pm$0.15 \\
18 & 111 & 60364.63551 & 2.51$\pm$0.20 & 2.82$\pm$0.26 & Y & 1.67 & 11.23$\pm$1.91 & 69.41$\pm$6.66 & 2.06$\pm$0.16 & 0.81$\pm$0.13 \\
19 & 114 & 60364.99047 & 2.47$\pm$0.22 & 2.87$\pm$0.28 & Y & 1.71 & 11.60$\pm$2.17 & 79.21$\pm$8.08 & 1.96$\pm$0.15 & 0.83$\pm$0.14 \\
20 & 114 & 60365.06182 & 2.48$\pm$0.21 & 2.86$\pm$0.29 & N & 1.71 & 11.55$\pm$2.12 & 73.67$\pm$7.75 & 1.90$\pm$0.14 & 0.82$\pm$0.14 \\
21 & 201 & 60365.13472 & 3.21$\pm$0.24 & 3.87$\pm$0.35 & N & 1.75 & 12.08$\pm$1.97 & 49.25$\pm$4.63 & 1.83$\pm$0.14 & 1.12$\pm$0.18 \\
22 & 202 & 60365.28943 & 2.27$\pm$0.22 & 2.73$\pm$0.28 & Y & 3.71/1.85 & 12.02$\pm$2.38 & 67.74$\pm$7.18 & 1.66$\pm$0.13 & 0.79$\pm$0.14 \\
23 & 204 & 60365.51993 & 2.33$\pm$0.12 & 2.80$\pm$0.29 & N & 5.53/1.84 & 12.05$\pm$1.87 & 60.53$\pm$6.60 & 1.50$\pm$0.11 & 0.81$\pm$0.14 \\
24 & 204 & 60365.60222 & 2.12$\pm$0.10 & 2.66$\pm$0.27 & N & 1.98 & 12.59$\pm$1.87 & 58.67$\pm$6.16 & 1.49$\pm$0.11 & 0.77$\pm$0.13 \\
25 & 205 & 60365.68368 & 2.32$\pm$0.12 & 2.86$\pm$0.27 & N & 1.96 & 12.32$\pm$1.81 & 61.06$\pm$6.08 & 1.49$\pm$0.11 & 0.83$\pm$0.14 \\
26 & 207 & 60366.01711 & 2.19$\pm$0.11 & 2.78$\pm$0.26 & N & 8.00/2.00 & 12.67$\pm$1.82 & 60.81$\pm$6.10 & 1.58$\pm$0.12 & 0.80$\pm$0.13 \\
27 & 209 & 60366.18713 & 2.30$\pm$0.11 & 3.14$\pm$0.30 & N & 4.08/2.04 & 13.67$\pm$1.96 & 69.09$\pm$7.00 & 1.63$\pm$0.12 & 0.90$\pm$0.15 \\
28 & 209 & 60366.27219 & 2.39$\pm$0.06 & 3.03$\pm$0.29 & Y & 2.04 & 12.66$\pm$1.57 & 74.42$\pm$7.60 & 1.66$\pm$0.13 & 0.87$\pm$0.15 \\
29 & 211 & 60366.43996 & 2.32$\pm$0.11 & 2.96$\pm$0.32 & N & 4.03/2.02 & 12.78$\pm$1.98 & 67.05$\pm$7.52 & 1.71$\pm$0.13 & 0.85$\pm$0.15 \\
30 & 211 & 60366.52374 & 2.41$\pm$0.14 & 3.00$\pm$0.30 & N & 2.01 & 12.43$\pm$1.94 & 71.81$\pm$7.45 & 1.73$\pm$0.13 & 0.86$\pm$0.15 \\
31 & 212 & 60366.61018 & 2.31$\pm$0.18 & 2.79$\pm$0.26 & N & 2.07 & 12.08$\pm$2.07 & 85.29$\pm$8.35 & 1.74$\pm$0.13 & 0.80$\pm$0.13 \\
32 & 214 & 60366.85996 & 2.67$\pm$0.07 & 3.11$\pm$0.30 & N & 2.03 & 11.68$\pm$1.42 & 70.15$\pm$7.18 & 1.77$\pm$0.13 & 0.90$\pm$0.15 \\
33 & 214 & 60366.94463 & 2.47$\pm$0.19 & 2.87$\pm$0.28 & N & 2.03 & 11.65$\pm$2.01 & 87.58$\pm$8.89 & 1.77$\pm$0.13 & 0.83$\pm$0.14 \\
34 & 215 & 60367.02904 & 2.34$\pm$0.12 & 2.90$\pm$0.28 & N & 2.03 & 12.40$\pm$1.87 & 88.92$\pm$9.13 & 1.77$\pm$0.13 & 0.84$\pm$0.14 \\
35 & 301 & 60367.11354 & $\--$ & $\--$ & $\--$ & 2.03 & $\--$ & $\--$ & $\--$ & $\--$ \\
36 & 302 & 60367.19894 & 2.45$\pm$0.05 & 2.97$\pm$0.29 & N & 2.05 & 12.14$\pm$1.45 & 74.33$\pm$7.56 & 1.77$\pm$0.13 & 0.86$\pm$0.14 \\
37 & 303 & 60367.37278 & 2.81$\pm$0.21 & 3.43$\pm$0.31 & N & 2.08 & 12.21$\pm$2.02 & 64.54$\pm$6.15 & 1.75$\pm$0.13 & 0.99$\pm$0.16 \\
38 & 303 & 60367.46190 & 2.61$\pm$0.08 & 3.45$\pm$0.36 & N & 2.14 & 13.22$\pm$1.77 & 69.11$\pm$7.52 & 1.73$\pm$0.13 & 1.00$\pm$0.17 \\
39 & 305 & 60367.64126 & 2.57$\pm$0.20 & 2.94$\pm$0.26 & N & 2.14 & 11.42$\pm$1.90 & 86.50$\pm$8.19 & 1.70$\pm$0.13 & 0.85$\pm$0.14 \\
40 & 305 & 60367.73144 & 2.60$\pm$0.06 & 3.10$\pm$0.27 & N & 2.16 & 11.96$\pm$1.33 & 82.58$\pm$7.71 & 1.68$\pm$0.13 & 0.89$\pm$0.14 \\
41 & 307 & 60367.91840 & 2.59$\pm$0.21 & 2.90$\pm$0.28 & Y & 2.28 & 11.23$\pm$2.02 & 89.34$\pm$9.17 & 1.63$\pm$0.12 & 0.84$\pm$0.14 \\
42 & 308 & 60368.01282 & 2.66$\pm$0.06 & 3.04$\pm$0.30 & N & 2.27 & 11.45$\pm$1.40 & 76.86$\pm$8.05 & 1.61$\pm$0.12 & 0.88$\pm$0.15 \\
43 & 309 & 60368.29469 & 2.69$\pm$0.20 & 3.12$\pm$0.29 & N & 2.27 & 11.59$\pm$1.95 & 63.25$\pm$6.23 & 1.53$\pm$0.12 & 0.90$\pm$0.15 \\
44 & 309 & 60368.39353 & 2.32$\pm$0.11 & 3.11$\pm$0.30 & N & 2.37 & 13.41$\pm$1.92 & 72.02$\pm$7.35 & 1.50$\pm$0.11 & 0.90$\pm$0.15 \\
45 & 309 & 60368.49190 & $\--$ & $\--$ & $\--$ & 2.36 & $\--$ & $\--$ & $\--$ & $\--$ \\
46 & 310 & 60368.58854 & 2.60$\pm$0.20 & 2.88$\pm$0.26 & N & 2.32 & 11.10$\pm$1.82 & 84.09$\pm$7.89 & 1.44$\pm$0.11 & 0.83$\pm$0.13 \\
47 & 311 & 60368.78858 & 2.38$\pm$0.12 & 2.97$\pm$0.29 & N & 2.56 & 12.47$\pm$1.81 & 63.50$\pm$6.42 & 1.38$\pm$0.10 & 0.86$\pm$0.14 \\
48 & 313 & 60369.00155 & 2.43$\pm$0.12 & 2.96$\pm$0.29 & N & 2.55 & 12.19$\pm$1.78 & 60.80$\pm$6.15 & 1.32$\pm$0.10 & 0.85$\pm$0.14 \\
49 & 314 & 60369.10782 & 2.42$\pm$0.06 & 2.88$\pm$0.27 & N & 2.55 & 11.91$\pm$1.38 & 54.62$\pm$5.31 & 1.29$\pm$0.10 & 0.83$\pm$0.14 \\
50 & 314 & 60369.21778 & 2.27$\pm$0.05 & 2.92$\pm$0.27 & Y & 2.64 & 12.90$\pm$1.50 & 56.05$\pm$5.48 & 1.26$\pm$0.10 & 0.84$\pm$0.14 \\
51 & 314 & 60369.33029 & 2.37$\pm$0.18 & 2.75$\pm$0.24 & Y & 2.70 & 11.57$\pm$1.87 & 60.37$\pm$5.55 & 1.22$\pm$0.09 & 0.79$\pm$0.13 \\
52 & 314 & 60369.44306 & 2.53$\pm$0.19 & 3.02$\pm$0.27 & Y & 2.71 & 11.92$\pm$1.96 & 47.79$\pm$4.54 & 1.19$\pm$0.09 & 0.87$\pm$0.14 \\
53 & 315 & 60369.55780 & 2.00$\pm$0.12 & 2.57$\pm$0.25 & Y & 2.75 & 12.85$\pm$2.02 & 51.21$\pm$5.31 & 1.16$\pm$0.09 & 0.74$\pm$0.13 \\
54 & 316 & 60369.79381 & 2.35$\pm$0.11 & 2.72$\pm$0.24 & N & 2.81 & 11.60$\pm$1.58 & 71.44$\pm$6.65 & 1.10$\pm$0.08 & 0.78$\pm$0.12 \\
55 & 317 & 60369.91172 & 2.30$\pm$0.10 & 2.71$\pm$0.24 & Y & 2.83 & 11.77$\pm$1.54 & 70.70$\pm$6.66 & 1.08$\pm$0.08 & 0.78$\pm$0.13 \\
56 & 401 & 60371.37052 & 2.44$\pm$0.20 & 2.57$\pm$0.23 & N & 3.29 & 10.53$\pm$1.79 & 55.99$\pm$5.27 & 0.89$\pm$0.07 & 0.74$\pm$0.12 \\
57 & 401 & 60371.50769 & 2.36$\pm$0.19 & 2.80$\pm$0.31 & N & 3.29 & 11.90$\pm$2.29 & 52.17$\pm$6.05 & 0.89$\pm$0.07 & 0.81$\pm$0.15 \\
58 & 501 & 60373.14843 & 2.75$\pm$0.27 & 3.10$\pm$0.29 & N & 3.47 & 11.27$\pm$2.18 & 102.55$\pm$10.15 & 1.43$\pm$0.11 & 0.89$\pm$0.15 \\
59 & 501 & 60373.29220 & 2.54$\pm$0.38 & 2.84$\pm$0.29 & N & 3.45 & 11.14$\pm$2.81 & 108.19$\pm$11.51 & 1.43$\pm$0.11 & 0.82$\pm$0.14 \\
60 & 501 & 60373.44431 & 2.60$\pm$0.11 & 2.86$\pm$0.25 & N & 3.65 & 11.01$\pm$1.42 & 113.48$\pm$10.47 & 1.43$\pm$0.11 & 0.82$\pm$0.13 \\
\enddata

\begin{tablenotes}
  \item $^{\rm a}$ The Obs. id. starts with P061437300. We only use the last three digits to represent \hxmt \ Obs. Ids.
  \item $^{\rm b}$ The observed recurrence time. If two values are given, the second one is obtained from the measured recurrence time divided by $N+1$, where $N$ is the number of missed bursts.
\end{tablenotes}
\end{deluxetable*}

\bibliography{sample631}{}
\bibliographystyle{aasjournal}

\end{document}